\PassOptionsToPackage{svgnames}{xcolor}
\documentclass[conference]{IEEEtran}
\usepackage{graphicx} 
\graphicspath{ {./share/} }
\IEEEoverridecommandlockouts
\usepackage{svg}
\usepackage{amsmath}
\usepackage{cite} 
\usepackage{booktabs}
\usepackage{multirow}
\usepackage{listings}
\usepackage{flafter}
\usepackage{booktabs}
\usepackage{array}
\usepackage{tikz}
\usepackage{xcolor}
\usepackage{hyperref}
\usepackage{siunitx}
\usepackage{microtype}

\title{Exploiting Music Source Separation for Automatic Lyrics Transcription with Whisper}

\newcommand\copyrighttext{%
  \footnotesize \textcopyright 2025 IEEE. Personal use of this material is permitted.
  Permission from IEEE must be obtained for all other uses, in any current or future
  media, including reprinting/republishing this material for advertising or promotional
  purposes, creating new collective works, for resale or redistribution to servers or
  lists, or reuse of any copyrighted component of this work in other works.
  }
\newcommand\copyrightnotice{%
\begin{tikzpicture}[remember picture,overlay]
\node[anchor=south,yshift=20pt] at (current page.south) {\fbox{\parbox{\dimexpr\textwidth-\fboxsep-\fboxrule\relax}{\copyrighttext}}};
\end{tikzpicture}%
}

\begin{document}

\author{\IEEEauthorblockN{
Jaza Syed\IEEEauthorrefmark{1}\IEEEauthorrefmark{3},
Ivan Meresman Higgs\IEEEauthorrefmark{1}, 
Ondřej Cífka\IEEEauthorrefmark{2} and
Mark Sandler\IEEEauthorrefmark{1}}
\IEEEauthorblockA{
\IEEEauthorrefmark{1}School of Electronic Engineering and Computer Science, Queen Mary University of London}
\IEEEauthorblockA{
\IEEEauthorrefmark{2}AudioShake}
\IEEEauthorblockA{
Email: j.syed@qmul.ac.uk,
i.meresman-higgs@qmul.ac.uk,
ondrej@audioshake.ai,
mark.sandler@qmul.ac.uk
}
}
\IEEEaftertitletext{
\thanks{
\IEEEauthorrefmark{3}Corresponding author
}
\thanks{
This work was supported by UKRI - InnovateUK [Grant Number 10102804]
}
}

\maketitle

\begin{abstract}
	Automatic lyrics transcription (ALT) remains a challenging task in the field of music
	information retrieval, despite great advances in automatic speech recognition (ASR)
	brought about by transformer-based architectures in recent years. One of the major
	challenges in ALT is the high amplitude of interfering audio signals relative to
	conventional ASR due to musical accompaniment. Recent advances in music source
	separation have enabled automatic extraction of high-quality separated vocals, which
	could potentially improve ALT performance. However, the effect of source separation has
	not been systematically investigated in order to establish best practices for its use.
	This work examines the impact of source separation on ALT using Whisper, a state-of-the-art
    open source ASR model. We evaluate Whisper's performance on original audio,
	separated vocals, and vocal stems across short-form and long-form transcription tasks.
	For short-form, we suggest a concatenation method that results in a consistent reduction in
	Word Error Rate (WER). For long-form, we propose an algorithm using source separation
	as a vocal activity detector to derive segment boundaries, which results in a
	consistent reduction in WER relative to Whisper's native long-form algorithm. Our
	approach achieves state-of-the-art results for an open source system on the Jam-ALT
	long-form ALT benchmark, without any training or fine-tuning. We also publish
	MUSDB-ALT, the first dataset of long-form lyric transcripts following the Jam-ALT
	guidelines for which vocal stems are publicly available.
\end{abstract}

\begin{IEEEkeywords}
	Automatic Lyrics Transcription, Voice Activity Detection, Source Separation
\end{IEEEkeywords}
\copyrightnotice

\section{Introduction}

Lyrics transcription is the task of obtaining lyrics from the singing voice in a
musical recording. Automatic lyrics transcription (ALT) systems enable lyrics
from large musical datasets to be obtained without requiring manual transcription. Recently, the open source
state-of-the-art (SOTA) automatic speech recognition (ASR) model Whisper
\cite{radford_robust_2023} has been found to perform well for ALT
\cite{cifka_lyrics_2024}, significantly outperforming domain-specific approaches
\cite{demirel_mstre-net_2021} and earlier ASR models fine-tuned for ALT
\cite{ou_transfer_2022}. LyricWhiz \cite{zhuo_lyricwhiz_2023} achieved SOTA performance
by combining multiple rounds of inference with Whisper with a closed source large language model
(LLM) as a text post-processor. Whisper outputs have been used for downstream music
information retrieval tasks that require lyrics transcription as a processing step
\cite{antonisen_polysinger_2024}, motivating the investigation of Whisper's
characteristics as an ALT system.

A key challenge in ALT is interference from musical accompaniment
\cite{fine_making_2014}, as vocal stems are not generally available. Both commercial
\cite{cifka_lyrics_2024} and open source \cite{ou_transfer_2022} ALT systems use music
source separation (MSS) to obtain separated vocals as a preprocessing step to reduce
interference. In \cite{ou_transfer_2022}, an ASR model is fine-tuned and evaluated on
vocals separated using Hybrid Demucs \cite{defossez_hybrid_2021}. However
\cite{cifka_lyrics_2024} finds that using separated vocals with Whisper for long-form
transcription can cause significant degradation in transcription quality. Vocal stems
with no musical accompaniment are expected to be easier to transcribe, however no
long-form ALT dataset with corresponding vocal stems is publicly available. As vocal
stems are the ultimate target of source separation, comparing performance with stems
and separated vocals may reveal areas for improving source separation for ALT.

Although Whisper has been successfully used for ALT, the model makes systematic errors in ALT
due to fundamental differences in the goals of ASR and ALT systems.
One such difference is the presence of ``backing vocals" in music, which are vocals
secondary to the lead vocal line. It is common practice to transcribe these, as they
may contribute to the lyrical content of the song, and to enclose them in parentheses
to distinguish them from the lead vocals \cite{cifka_lyrics_2024}. In contrast, in ASR
it is often desirable for a model to ignore background speech~-- Whisper is evaluated
on robustness to additive ``pub noise" \cite{radford_robust_2023} which may contain
speech. Song lyrics also contain non-lexical vocables \cite{chambers1980nonlexical}
such as \emph{ooh}, \emph{ah}, and
\emph{la}. Recent work
\cite{romana_automatic_2024} has shown that such vocalizations are often discarded by
ASR models such as Whisper.

In ALT datasets such as JamendoLyrics MultiLang
\cite{stoller_end--end_2019} and DALI \cite{meseguer-brocal_dali_2018}, non-lexical
vocables are often inconsistently annotated, using variable numbers of vowels to
emphasize syllable length or with inconsistent numbers of syllables of
multi-syllable vocalizations.  wh
Jam-ALT \cite{cifka_lyrics_2024} is the first dataset for long-form ALT 
with consistently transcribed backing vocals and
non-lexical vocables. Jam-ALT also has
accompanying annotation guidelines that align with industry standards for lyrics
transcription and formatting. Lyrics annotations \cite{schulze-forster_phoneme_2021}
have been published for the widely used MSS dataset MUSDB18
\cite{rafii_musdb18-hq_2019}, however they are only provided for short
segments which may overlap in time, so cannot be used for long-form transcription. They also do
not follow the Jam-ALT guidelines for backing vocals and non-lexical vocables.

In addition, ALT systems often process extended recordings exceeding their underlying
model's input length. While the ASR and ALT literature uses the terms ``short-form" and
``long-form" inconsistently, in Whisper's context, these terms have specific meanings:
the underlying neural network accepts audio segments under \SI{30}{s} (``short-form"),
requiring segmentation strategies for longer recordings (``long-form").  Whisper's
performance on long-form transcription tasks is sensitive to the segmentation algorithm used
\cite{radford_robust_2023}. WhisperX \cite{bain_whisperx_2023} improves on the
performance of Whisper's native long-form segmentation algorithm for long-form ASR
transcription tasks by using Voice Activity Detection (VAD) to determine segments, but
has not been evaluated for long-form ALT. 

In short-form ALT, the effect of
MSS on Whisper's output can be observed independently of potential interactions with the long-form
transcription algorithm, which may affect the sections of audio transcribed. Since short-form ALT is a sub-task of long-form, findings should be transferable, however the two have not been systematically compared.

In this work, we study the effect of MSS on both short-form and long-form ALT. In order to determine the effect of MSS artifacts relative to
``perfect" source separation, we construct a new dataset \textbf{MUSDB-ALT} of
long-form lyrics transcripts of a subset of the MUSDB18 test set, following the Jam-ALT
guidelines. We evaluate Whisper's performance on Jam-ALT and MUSDB-ALT using the
original audio, separated vocals for both datasets, and vocal stems for MUSDB-ALT. Our experimental code is published open source\footnote{\url{https://github.com/jaza-syed/mss-alt}}. The
contributions of our work are:

\begin{itemize}
	\item We create and publish MUSDB-ALT\footnote{\url{https://huggingface.co/datasets/jazasyed/musdb-alt}}, the first consistently annotated dataset of lyrics for long-form ALT for
	      which vocal stems are publicly available
	\item We publish line timings and annotations for non-lexical vocables in
	      Jam-ALT\footnote{\url{https://huggingface.co/datasets/audioshake/jam-alt}} and
	      MUSDB-ALT
	\item We show that short-form performance is strongly affected by sample length and propose
	      a method of merging samples so that performance is similar to long-form
	\item We propose using separated vocals to obtain segment boundaries for
	      long-form transcription, achieving SOTA performance on Jam-ALT for an open source
	      system
\end{itemize}

\section{Methods}
In the following sections, we introduce our evaluation datasets, preprocessing methods for audio and
lyrics, metrics, and specific experimental methods for short-form and long-form transcription.

\subsection{Data}
\label{section:data}
{\renewcommand{\arraystretch}{1.1}
	\addtolength{\tabcolsep}{-0.2em}
	\begin{table}[t!]
		\centering
		\caption{Summary of datasets}
		\label{tab:summary}
		\begin{tabular}{c|cccc}
    \toprule
    \textbf{Dataset} & \textbf{Songs} & \textbf{Duration (min)} & \textbf{Non-lexical\%} & \textbf{Backing\%} \\
    \midrule
    Jam-ALT          & 79             & 283                      & 4.64                   & 4.66               \\
    MUSDB-ALT        & 39             & 166                      & 3.27                   & 4.90               \\
    \bottomrule
\end{tabular}
	\end{table}
	\renewcommand{\arraystretch}{1 }
}
Our work uses Jam-ALT \cite{cifka_lyrics_2024}, an updated version of the JamendoLyrics
MultiLang \cite{stoller_end--end_2019} benchmark. Jam-ALT contains 79 songs in English,
French, Spanish and German with full-song transcripts.
To enable using Jam-ALT for short-form transcription, we enriched it with line-level
timings based on JamendoLyrics. Specifically, we computed the optimal word alignment
(edit-distance-based, as explained below in Section \ref{section:metrics}) between the
JamendoLyrics and Jam-ALT transcripts, and used this alignment, along with the
word-level timings in JamendoLyrics, to automatically derive line timings for Jam-ALT.
We subsequently manually revised the timings to account for ambiguity in the automatic
assignment (often caused by missing words or lines in JamendoLyrics), as well as
inaccurate timing annotations in JamendoLyrics.

As another evaluation set, we propose MUSDB-ALT, a novel dataset of long-form
transcripts with line-level timings following the Jam-ALT guidelines. We produced
MUSDB-ALT manually based on the MUSDB lyrics extension
\cite{schulze-forster_phoneme_2021}. The major changes we made were ensuring that
non-lexical vocables were consistently transcribed, backing vocals were enclosed in
parentheses and that line and section breaks in the transcripts corresponded to
musically significant sections.

MUSDB-ALT consists of 39 of the 45 English-language songs in the MUSDB18 test set. Of
the six songs excluded from the dataset, three contained primarily instrumental music
with minimal lyrical content derived from highly processed vocal samples, two featured
extended sections with three or more interacting vocal lines that could not be
categorized as lead and backing vocals, and one was omitted due to screamed vocals with
unintelligible lyrical content. We use audio from MUSDB18-HQ
\cite{rafii_musdb18-hq_2019}, the uncompressed version of MUSDB18.

In both datasets backing vocals are identifiable as they are within parentheses. We
manually produced extra annotations of which words are non-lexical vocables. We use
these annotations to define metrics specific to these words in Section
\ref{section:metrics-deletions}. The dataset duration, number of songs and percentage
of words in the data that are non-lexical vocables or backing vocals are shown in Table
\ref{tab:summary}.

\subsection{Metrics}
\label{section:metrics}
\begin{figure}[!t]
	\centering
        \vspace{-0.9em}
	\begin{tikzpicture}
		\draw[thick, fill=gray!10] (0,0) rectangle (8, -1.4);
		\node[anchor=north west, text width=7.6cm, font=\ttfamily] at (0.2, -0.2) {\small
			\obeyspaces
			I \color{blue} went \color{black} to the \color{DarkGreen} +++ \color{blue}park \color{black} yesterday \color{red} evening  \color{black}\\
			I \color{blue} came \color{black}to the \color{DarkGreen} new \color{blue} pool \color{black} yesterday \color{red} -------
		};
	\end{tikzpicture}
	\caption{Example of an alignment with reference above and hypothesis below.
		Substitutions are in {\color{blue}blue}, hits are in black, insertions
		are denoted by \texttt{+} in {\color{DarkGreen}green} and are deletions
		denoted by \texttt{-} in {\color{red}red}.}
	\label{fig:alignment}
\end{figure}

The standard metric to evaluate ASR performance is the Word Error Rate (WER),
which is calculated as the normalized minimum edit distance (Levenshtein distance) between what an ALT system
produces (the hypothesis) and the ground-truth transcript (the reference). WER
considers the minimum number of word-level edit operations (deletions, insertions, and substitutions) required to transform the
reference into the hypothesis. Denoting the numbers of deletions
as $D$, substitutions as $S$, insertions as $I$, and the number of
correctly recognized words (hits) as $H$,
the WER is defined as:
\begin{equation}
	\mathit{WER} = \frac{S + D + I}{S + D + H}
\end{equation}
Defining the total length of the reference $N=S+D+H$ for brevity, we can disaggregate the WER into
	$\mathit{SR}  = S/N$, 
	$\mathit{DR}  =D/N$ and
	$\mathit{IR} =I/N$  for substitutions, deletions and insertions respectively.

When computing the WER, we apply lyrics-specific punctuation normalization and
tokenization from the
\texttt{alt-eval}\footnote{\url{https://github.com/audioshake/alt-eval}}
package to both the reference and hypothesis, and remove all non-word tokens, following
\cite{cifka_lyrics_2024}. We compute the WER for multiple songs or short-form samples
by summing the component edit counts $H$, $D$, $S$, and $I$ over all the songs or samples.
The minimal sequence of edits produced by the edit distance calculation can be
interpreted as an \emph{alignment}, where each hit or substitution aligns a hypothesis
word to a reference word; insertions and deletions correspond to unpaired words in the
hypothesis and the reference respectively. An example of an alignment is shown in
Figure \ref{fig:alignment}. Using the alignment, we define additional rates presented in the sections below.

\subsubsection{Hallucination rate}
Whisper can produce hallucinations - text totally unrelated to the audio input or
repetition loops of the same short text segment \cite{ye_spurious_2024}. As they are
unrelated to the audio, these hallucinations frequently appear in the alignment as
sequential insertions. We define $\mathit{I}_{\text{10}}$ as the number of insertions
within blocks of 10 or more sequential insertions, and define a proxy hallucination
rate $\mathit{IR}_{\text{10}} = {I_{10}} / {N}$.
Manual inspection confirmed these blocks are hallucinations with a low false positive rate, 
though the method fails with short hallucinations or when chance matches disrupt insertions.

{\addtolength{\tabcolsep}{-0.23em}
\begin{table}[t!]
	\centering
	\caption{Summary statistics of merged and grouped lines}
	\label{tab:segment_length}
	\begin{tabular}{ll|rrrr|r}
	\toprule
	                              &               & \multicolumn{4}{|c|}{\textbf{Duration (s)}}                                                                        \\
	\cmidrule(lr){3-6}
	\textbf{Dataset}              & \textbf{Type} & \textbf{Mean}                               & \textbf{Std. Dev.} & \textbf{ Min.} & \textbf{Max.} & \textbf{Count} \\
	\midrule
	\multirow[c]{2}{*}{Jam-ALT}   & Merged Line   & 3.52                                        & 1.80               & 0.44           & 17.66         & 3445           \\
	                              & Group         & 20.44                                       & 5.51               & 3.64           & 29.93         & 613            \\
	\multirow[c]{2}{*}{MUSDB-ALT} & Merged Line   & 4.59                                        & 2.39               & 1.22           & 23.82         & 1488           \\
	                              & Group         & 19.84                                       & 5.51               & 2.27           & 29.93         & 359            \\
	\bottomrule
\end{tabular}
\end{table}
}

\subsubsection{Deletions}
\label{section:metrics-deletions}
The most common cause of error for non-lexical vocables and backing vocals is deletions. 
Counting the number of deletions where the reference words are non-lexical vocables
$D_{\mathit{NL}}$ and backing vocals $D_{\mathit{BV}}$, we define 
	$\mathit{DR}_{\mathit{NL}}  = {D_{\mathit{NL}}}/{N}$ and 
	$\mathit{DR}_{\mathit{BV}}  = {D_{\mathit{BV}}}/{N}$
 to disaggregate the deletion rate by reference word type.

\subsection{Source separation}
We use two Hybrid Demucs \cite{defossez_hybrid_2021} pre-trained models: \texttt{mdx}
(trained on the MUSDB18 train set only) and \texttt{mdx\_extra} (trained on the MUSDB18
train and test sets plus 800 additional songs). These models achieved Signal-to-Distortion
Ratios (SDR) for vocals of \SI{7.97}{\decibel} and \SI{8.76}{\decibel} respectively on the MUSDB18 test set,
demonstrating that \texttt{mdx\_extra} improves the quality of separated vocals over
\texttt{mdx}. When using \texttt{mdx\_extra} for source separation with MUSDB-ALT, we
are evaluating on the training set, meaning we likely overestimate performance on
unseen data.

{\addtolength{\tabcolsep}{-0.3em}
    \begin{table*}[t!]
        \centering
        \caption{Short-form results}
        \label{tab:short_results}
        \begin{tabular}{ll|rrrrrrr|rrrrrrr}
    \toprule
    \multicolumn{2}{c|}{} & \multicolumn{7}{c|}{\textbf{Jam-ALT}} & \multicolumn{7}{c}{\textbf{MUSDB-ALT}}                                                                                                                                                                                                                                                                                                                                                                                                  \\
    \cmidrule(lr){3-9} \cmidrule(lr){10-16}
    \textbf{Type}         & \textbf{Audio}                        & $\mathit{WER}$                         & \textbf{$\mathit{SR}$} & \textbf{$\mathit{DR}$} & \textbf{$\mathit{IR}$} & $\mathit{IR}_\text{10}$ & \textbf{$\mathit{DR}_\mathit{NL}$} & \textbf{$\mathit{DR}_\mathit{BV}$} & $\mathit{WER}$ & \textbf{$\mathit{SR}$} & \textbf{$\mathit{DR}$} & \textbf{$\mathit{IR}$} & $\mathit{IR}_\text{10}$ & \textbf{$\mathit{DR}_\mathit{NL}$} & \textbf{$\mathit{DR}_\mathit{BV}$} \\
    \midrule
    \multirow{4}{*}{Group}
                          & Original Mix                          & \textbf{20.99}                         & 9.69                   & 9.54                   & \textbf{1.76}          & 0.05                             & 3.54                               & 3.07                               & 23.59          & 9.24                   & 12.31                  & 2.03                   & \textbf{0.12}                    & 1.77                               & 3.31                               \\
                          & Separated (\texttt{mdx})              & 21.17                                  & 9.25                   & 10.14                  & 1.78                   & \textbf{0.00}                    & 3.40                               & 3.12                               & 23.98          & 9.56                   & 10.47                  & 3.96                   & 1.80                             & 1.68                               & 3.03                               \\
                          & Separated (\texttt{mdx\_extra})       & 21.08                                  & \textbf{9.02}          & \textbf{9.04}          & 3.01                   & 1.29                             & \textbf{3.24}                      & \textbf{3.04}                      & \textbf{20.00} & \textbf{8.16}          & \textbf{9.81}          & \textbf{2.03}          & \textbf{0.12}                    & \textbf{1.64}                      & \textbf{2.77}                      \\
                          & Vocal Stem                            & -                                      & -                      & -                      & -                      & -                                & -                                  & -                                  & 14.19          & 4.94                   & 7.75                   & 1.51                   & 0.21                             & 1.51                               & 2.90                               \\
    \midrule
    \multirow{4}{*}{Merged Line}
                          & Original Mix                          & 29.46                                  & 17.86                  & \textbf{7.67}          & 3.93                   & \textbf{0.00}                    & 1.95                               & 2.89                               & 28.57          & 15.17                  & 10.22                  & \textbf{3.18}          & \textbf{0.00}                    & \textbf{1.40}                      & 2.96                               \\
                          & Separated (\texttt{mdx})              & 29.88                                  & 17.69                  & 8.13                   & 4.07                   & 0.46                             & \textbf{1.90}                      & \textbf{2.84}                      & 29.89          & 14.65                  & 9.40                   & 5.84                   & 2.63                             & \textbf{1.40}                      & 2.76                               \\
                          & Separated (\texttt{mdx\_extra})       & \textbf{28.06}                         & \textbf{16.53}         & 7.88                   & \textbf{3.66}          & \textbf{0.00}                    & 1.96                               & 2.88                               & \textbf{25.40} & \textbf{13.30}         & \textbf{8.10}          & 4.00                   & 1.39                             & 1.44                               & \textbf{2.60}                      \\
                          & Vocal Stem                            & -                                      & -                      & -                      & -                      & -                                & -                                  & -                                  & 14.93          & 6.88                   & 6.27                   & 1.78                   & 0.00                             & 1.22                               & 2.63                               \\
    \bottomrule
\end{tabular}  
    \end{table*}}

\subsection {Whisper}
We use the \emph{Faster
	Whisper}\footnote{\url{https://github.com/SYSTRAN/faster-whisper}}
implementation of Whisper. \emph{Faster Whisper} features batched transcription over
segments of the same audio file, which we use for speed where possible. We use Whisper model
\texttt{large-v2} and provide the language of the song at transcription time, following
\cite{cifka_lyrics_2024}. For decoding, we use a beam size of 5. We compute average
error rates over 5 runs to account for Whisper's stochastic decoding algorithm.

\subsection{Short-form transcription}
Some lyric lines may overlap in time, so lines cannot directly be used as samples for
short-form transcription. We use two methods to merge lines to produce non-overlapping samples of different average lengths, in order to investigate the effect of segment
length on ALT performance.

\subsubsection {Merging}
\label{section:methods-short-merge}
Using a threshold of \SI{0.2}{s} overlap between lines, we identify sets of transitively overlapping lines. To define a new short-form sample for each set, we concatenate the lines to obtain a transcript and use the earliest line start time and latest line end time as the timings. We exclude any samples longer than \SI{30}{s}, of which there are
none in Jam-ALT and only one in MUSDB-ALT. We refer to these samples as ``merged
lines". Since this process excludes so few lyrics,
metrics for short-form and long-form transcription are comparable.

\subsubsection {Groups}
To produce a second set of short-form samples of greater average duration, we apply an additional
two-step grouping procedure to the merged lines for each song:
\begin{itemize}
	\item We split the lines into groups on gaps of longer than \SI{7}{s}. between the end and start of
	      consecutive lines.
	\item We split the groups into one or more subgroups of consecutive lines such that the
	      minimum subgroup duration is maximized and the maximum is under \SI{30}{s}.
\end{itemize}
Each group of merged lines produces a short-form sample, where the earliest start time and latest end time are the timings of the new sample and the concatenated transcripts are the new transcript. The durations of the merged lines and groups are summarized in Table \ref{tab:segment_length}.

\subsection{Long-form transcription}
In the following section, we present a novel segmentation algorithm for long-form ALT
and describe considerations specific to our use of Whisper's native long-form
algorithm.
\subsubsection{RMS-VAD}
Applying a threshold to the root mean square (RMS) amplitude of separated vocals is an
effective VAD for singing voice \cite{bonzi_exploiting_2023}. For a signal $x[n]$ where
$n$ is the sample index, we compute the time-varying RMS amplitude with frame size $N$
as
\begin{equation}
	\text{RMS}[n] = \sqrt{\frac{1}{N}\sum_{m=n-N+1}^{n} x[m]^2}.
\end{equation}
We then define the normalised score $\text{VAD}[n]$ as
\begin{equation}
	\text{VAD}[n] = \frac{\text{RMS}[n]}{\max_{m} \text{RMS}[m]}.
\end{equation}
The ``Cut \& Merge" segmentation algorithm defined in \cite{bain_whisperx_2023}
converts VAD scores into non-overlapping vocal activity segments optimized to approach
a specified maximum length. Whisper's output for each segment can then be concatenated
to obtain the final transcript. We extend the approach in \cite{bonzi_exploiting_2023} to define a new segmentation
algorithm \textbf{RMS-VAD}:
\begin{enumerate}
	\item Obtain separated vocals with a source separation model
	\item Compute $\text{VAD}[n]$ for the separated vocals
	\item Obtain boundaries from $\text{VAD}[n]$ using ``Cut \& Merge"
\end{enumerate}

For ``Cut \& Merge", we use onset and offset thresholds of 0.1, a minimum silence
duration of \SI{1}{s} and a maximum segment length of \SI{30}{s}, matching Whisper's input
duration.

\begin{figure}[b!]
    \vspace{-1em}
	\includegraphics[width=0.96\linewidth]{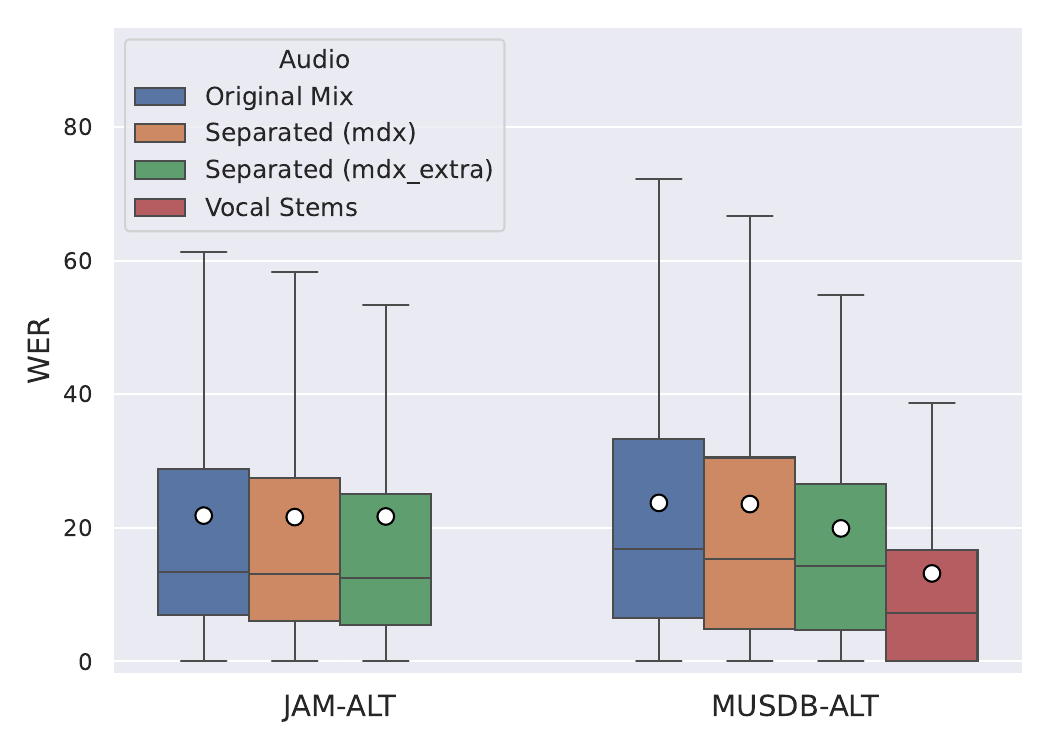}
	\caption{Group-level WER distribution by dataset and audio. Means are shown by white
		circles. Outliers which affect the means occur but are not shown. WER medians and quartiles improve slightly with separation quality for JAM-ALT and more significantly for MUSDB-ALT.}
	\label{fig:short_box}
\end{figure}

\subsubsection {Whisper}
Whisper groups its text output into sections with predicted start and end timestamps. 
Its native long-form transcription takes the last predicted timestamp in a \SI{30}{s} window 
as the start of the next
window. This approach prevents batched transcription of sections of a
single song unlike when segment timings are provided by algorithms such as RMS-VAD. In long-form
transcription, it is possible to use Whisper's prompt to condition the output for a
segment on the text predicted for the previous segment. Since this conditioning is
incompatible with batched transcription, we disabled it in the native long-form
algorithm. This change had no significant impact on performance.

\section{Experiments}
\label{section:results}
In the following sections, we present the results of our experiments on both short-form and
long-form transcription tasks. In both experiments, we report the metrics described in Section \ref{section:metrics} as percentages.
\subsection{Short-form transcription}
We evaluate Whisper's short-form performance, using both the merged lines and groups
described in Section \ref{section:data} as samples. For Jam-ALT and MUSDB-ALT, we evaluate with the original
audio, vocals separated with \texttt{mdx} and vocals separated with \texttt{mdx\_extra}. We also evaluate on vocal stems for MUSDB-ALT. We show the
results in Table \ref{tab:short_results} with the lowest error rates in bold, except for
vocal stems.
	{\addtolength{\tabcolsep}{-0.2em}
		\begin{table*}[t!]
			\centering
			\caption{Long-form results}
			\label{tab:long_results}
			\begin{tabular}{ll|rrrrrrr|rrrrrrr}
    \toprule
                   &                    & \multicolumn{7}{c|}{\textbf{Jam-ALT}} & \multicolumn{7}{c}{\textbf{MUSDB-ALT}}                                                                                                                                                                                                                                                                                                                                                                         \\
    \cmidrule(lr){3-9} \cmidrule(lr){10-16}
    \textbf{Audio} & \textbf{Algorithm} & $\mathit{WER}$                        & \textbf{$\mathit{SR}$}                 & \textbf{$\mathit{DR}$} & \textbf{$\mathit{IR}$} & $\mathit{IR}_\text{10}$ & \textbf{$\mathit{DR}_\mathit{NL}$} & \textbf{$\mathit{DR}_\mathit{BV}$} & $\mathit{WER}$ & \textbf{$\mathit{SR}$} & \textbf{$\mathit{DR}$} & \textbf{$\mathit{IR}$} & $\mathit{IR}_\text{10}$ & \textbf{$\mathit{DR}_\mathit{NL}$} & \textbf{$\mathit{DR}_\mathit{BV}$} \\
    \midrule
    \multirow[c]{2}{*}{Original Mix}
                   & Native             & 23.02                                 & 10.84                                  & 9.16                   & 3.03                   & 1.31                             & 3.29                               & \textbf{2.69}                      & 25.82          & 10.39                  & 12.25                  & 3.19                   & 1.09                             & 1.97                               & 3.31                               \\
                   & RMS-VAD            & \textbf{20.35}                        & 9.94                                   & \textbf{8.74}          & 1.67                   & \textbf{0.35}                    & 3.25                               & 3.05                               & 22.72          & 9.24                   & 11.01                  & 2.47                   & \textbf{0.12}                    & 1.89                               & 3.30                               \\
    \multirow[c]{2}{*}{Separated Vocals}
                   & Native             & 22.87                                 & 9.64                                   & 10.06                  & 3.18                   & 1.35                             & \textbf{3.21}                      & 2.88                               & 21.90          & 9.07                   & \textbf{9.71}          & 3.12                   & 0.80                             & \textbf{1.47}                      & \textbf{2.92}                      \\
                   & RMS-VAD            & 20.72                                 & \textbf{8.65}                          & 10.48                  & \textbf{1.58}          & 0.41                             & 3.32                               & 3.04                               & \textbf{20.07} & \textbf{7.87}          & 10.14                  & \textbf{2.06}          & 0.24                             & 1.93                               & 3.08                               \\
    \multirow[c]{2}{*}{Vocal Stem}
                   & Native             & -                                     & -                                      & -                      & -                      & -                                & -                                  & -                                  & 17.51          & 6.31                   & 8.26                   & 2.93                   & 0.95                             & 1.54                               & 2.87                               \\
                   & RMS-VAD            & -                                     & -                                      & -                      & -                      & -                                & -                                  & -                                  & 14.98          & 4.85                   & 8.53                   & 1.60                   & 0.12                             & 1.89                               & 3.09                               \\
    \bottomrule
\end{tabular}
		\end{table*}
	}

\newcommand*{\drnl}{\ensuremath{\mathit{DR}_\mathit{NL}}}
\newcommand*{\drbv}{\ensuremath{\mathit{DR}_\mathit{BV}}}
\newcommand*{\irhal}{\ensuremath{\mathit{IR}_\text{10}}}
\subsubsection{Sample length}
We observe that evaluating on groups yields lower WER than evaluating on merged lines
in almost all cases. This improvement stems from reduced substitutions, highlighting the importance of extended audio context to increase the
intelligibility of the vocals to Whisper. Deletions are mostly lower for merged lines
than for groups, except for backing vocals. Interestingly, the lower
deletions are mostly accounted for by a reduction in $\drnl$ for Jam-ALT. Group and line
evaluations show similar performance only for MUSDB-ALT stems, suggesting that musical
accompaniment or source separation artifacts are what cause sample durations to affect
performance significantly.

Error rates for groups are similar to those for long-form transcription in Table
\ref{tab:long_results}, whereas line-level performance is significantly worse. This is
likely because the segment lengths in long-form transcription are close to \SI{30}{s}.
As longer sample length is crucial to Whisper's short-form ALT performance, we only consider
group-level evaluation in the remaining analysis.

\subsubsection{Source separation}
For Jam-ALT the WER remains essentially unchanged for both original and separated audio. 
However, the WER for \texttt{mdx\_extra} is skewed higher due to the presence of a significant amount of hallucinations. 
$\mathit{SR}$ is improved by MSS and is better for \texttt{mdx\_extra} than \texttt{mdx}, demonstrating that increasing MSS quality improves vocal intelligibility for Jam-ALT.
Subtracting $\irhal$, the same trend applies to the WER. 
Source separation has little impact on reducing deletions for Jam-ALT and shows negligible effects on insertions, especially when hallucinations are excluded.

For MUSDB-ALT, the overall WER and components $\mathit{SR}$, $\mathit{DR}$ and $\mathit{IR}$ consistently decrease as separation quality improves.  
Notably, the \texttt{mdx\_extra} model, which was trained on MUSDB-ALT songs, shows a substantial improvement. Deletions are significantly lower for stems than for separated vocals, indicating that MSS artifacts cause Whisper to fail to transcribe some vocals.

Figure \ref{fig:short_box} shows that the mean and quartiles of the WER distribution follow the same trend for both datasets, with reduced effect for Jam-ALT.
Improved separation quality reduces substitutions across both datasets, indicating that source separation increases vocal intelligibility for Whisper.
\subsubsection{Hallucinations}
A low proportion of the WER is accounted for by $\irhal$, particularly for Jam-ALT.
$\irhal$ is consistently low for the original audio and vocal stems but is
higher for some cases with MSS, indicating that source separation artifacts can trigger
hallucinations.
\begin{figure}[b!]
   \vspace{-1em}
	\includegraphics[width=\linewidth]{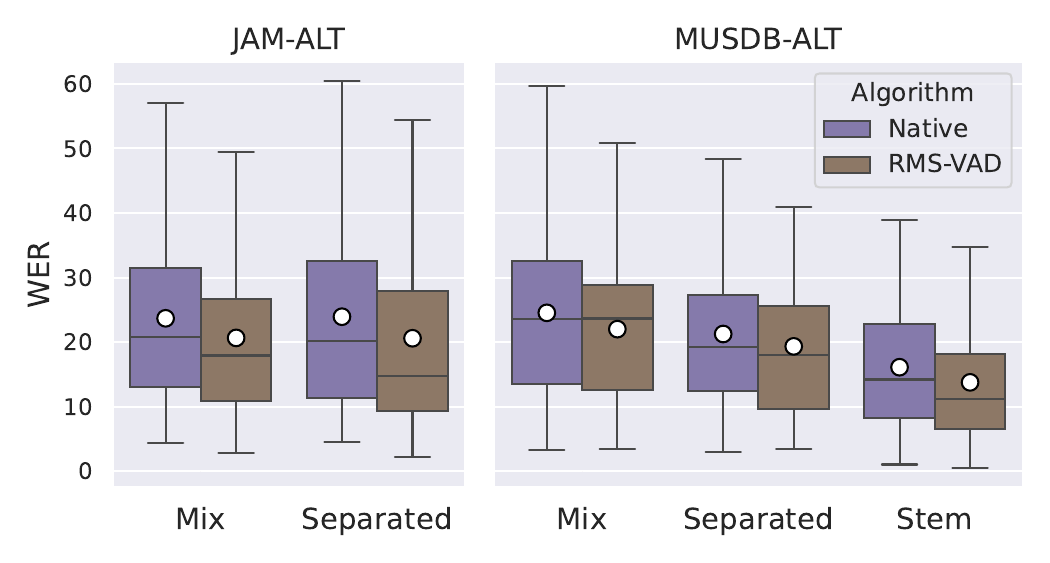}
	\caption{Song-level WER distribution for each dataset, by audio and long-form
		transcription algorithm used.  Means are shown by white circles. Outliers
		which affect the means occur but are not shown. Medians and quartiles of the WER are improved by RMS-VAD for all audio types and datasets.}
	\label{fig:long_box}
\end{figure}
\subsubsection{Deletions}
Non-lexical vocables and backing vocals constitute a high proportion of total
deletions. Comparing their occurrence rates in Table \ref{tab:summary} with
$\mathit{DR}_\mathit{NL}$ and $\mathit{DR}_\mathit{BV}$ in Table
\ref{tab:short_results} shows that over half of both are deleted in both datasets, revealing a shortcoming in Whisper. For Jam-ALT, MSS does not impact these
rates. While improved separation quality reduces both rates for MUSDB-ALT, the proportionate decrease is significantly less than
the proportionate decrease in $\mathit{DR}$. The high values
observed even with vocal stems indicate that this issue cannot be resolved by
improved source separation.

\subsection{Long-form transcription}
For long-form transcription, we use only \texttt{mdx\_extra} for source separation
because it demonstrated superior performance for short-form transcription. We evaluate
on original mixes, separated vocals and vocal stems using two segmentation approaches:
(1) Whisper's native long-form transcription algorithm, which determines boundaries
by predicting timestamps in the audio being transcribed and (2) our approach RMS-VAD, which determines
boundaries using the amplitude of separated vocals. When evaluating with RMS-VAD, we use the boundaries
obtained from the separated vocals even when evaluating on original mixes and vocal
stems. This design allows us to isolate the effect of RMS-VAD boundaries from the
effect of the audio type. We show the results in Table \ref{tab:long_results} with the lowest error rates in bold,
except for vocal stems.

\begin{table}[t!]
	\centering
	\caption{Long-form WER for Jam-ALT by language}
	\label{tab:sota}
	\begin{tabular}{l *{5}{>{\centering\arraybackslash}c}}
    \toprule
                                         &                    & \multicolumn{4}{c}{\textbf{Language}}                                                    \\
    \cmidrule(lr){3-6}
    \textbf{Audio}                       & \textbf{Algorithm} & DE                                    & EN             & ES             & FR             \\
    \midrule
    \multirow[c]{2}{*}{Original Mix}     & Native             & 18.14                                 & 26.92          & 17.07          & 24.98          \\
                                         & RMS-VAD            & 16.12                                 & \textbf{24.73} & \textbf{14.28} & \textbf{24.41} \\
    \cmidrule(lr){1-6}
    \multirow[c]{2}{*}{Separated Vocals} & Native             & 15.75                                 & 25.41          & 19.80          & 25.68          \\
                                         & RMS-VAD            & \textbf{13.03}                        & \textbf{24.73} & 17.86          & 24.50          \\
    \cmidrule(lr){1-6}
                                &   LyricWhiz                 & -                                     & 23.70          & -              & -              \\
    \bottomrule
\end{tabular}

\end{table}

\subsubsection{Effect of RMS-VAD}
For Jam-ALT, source separation provides little improvement when used with the native
algorithm, which is consistent with the short-form Jam-ALT results that indicated
little improvement. Using RMS-VAD improves the WER over Whisper's native
algorithm for both the original mix and the separated vocals. RMS-VAD shows similar
performance with original and separated audio, indicating that the improvement comes
from better segment boundaries rather than improved vocal quality after separation.

For MUSDB-ALT, RMS-VAD demonstrates consistent improvements over the native algorithm
on original mixes, separated vocals and even stems, which highlights the consistency of
the segmentation approach in enhancing transcription accuracy. Figure
\ref{fig:long_box} shows that the quartiles and medians of the song-level WER
distribution are also improved by RMS-VAD.

\subsubsection{Hallucinations and deletions}
The use of RMS-VAD has no significant effect on $\drnl$ or $\drbv$ however it does
reduce $\irhal$ relative to the native algorithm for all audio types. With
RMS-VAD, $\irhal$ values are similar to those for short-form transcription with the
same audio, with the exception of \texttt{mdx\_extra} on JAM-ALT. Whisper's native algorithm uses its predicted timestamps
to avoid cutting speech with segment boundaries \cite{radford_robust_2023}, but these timestamps can be inaccurate
\cite{zusag24_interspeech}. In contrast, short-form transcription uses gold-standard
human-annotated line boundaries and RMS-VAD is unlikely to cut words as it uses a low
vocal amplitude threshold. Therefore, we hypothesize that segment boundaries
interrupting vocal activity can cause hallucinations.

\subsubsection{Comparison with SOTA}
We report the long-form WER by song language for Jam-ALT in Table \ref{tab:sota}. We include the WER for LyricWhiz, which is only
evaluated on the original mixes of the English songs, for comparison. The previous open source SOTA for Jam-ALT was Whisper using its native
algorithm with provided language \cite{cifka_lyrics_2024}, which corresponds to the top row of both and Table \ref{tab:long_results} and \ref{tab:sota}. RMS-VAD boundaries improve the WER relative to the
native algorithm for all languages on both original and separated audio. RMS-VAD evaluated on original mixes achieves a new open source SOTA WER of 20.35. For English, RMS-VAD
provides an improvement on Whisper close to the effect of LyricWhiz without the use of multiple inference rounds or an LLM.

\section{Conclusion}
In this work, we show that Whisper's performance on line-level short-form transcription
(under \SI{30}{s}) is not indicative of performance on long-form transcription (over
\SI{30}{s}) for ALT due to the short duration of lines.

We introduce a method for grouping lines into longer short-form samples, which improves the utility of short-form evaluation 
for assessing the impact of audio preprocessing methods such as source separation on downstream performance.
For long-form transcription, we introduce a new method RMS-VAD to obtain segment
boundaries using source separation, which yields consistent improvements over 
Whisper's native long-form algorithm. In practice, an ALT system may be run over large
datasets of hundreds of thousands of songs. As source separation is computationally
expensive, developing less resource-intensive music-specific VAD systems to obtain
segment boundaries directly may therefore be a valuable direction for future work.
In addition, future work on ALT systems would benefit from considering the importance 
of sample length and segment boundary placement demonstrated in this study.

We also show that Whisper
systematically deletes non-lexical vocables and backing vocals, regardless of source
separation quality. These deletions constitute a high proportion of the total, indicating that 
this is an area for significant potential improvement.

\section*{Acknowledgments}
We thank Tom Andersson and Yasmin Gapper for feedback on the manuscript.

\vspace{-0.2em}

\bibliographystyle{IEEEtran} \bibliography{references_final}

\begin{thebibliography}{10}
\providecommand{\url}[1]{#1}
\csname url@samestyle\endcsname
\providecommand{\newblock}{\relax}
\providecommand{\bibinfo}[2]{#2}
\providecommand{\BIBentrySTDinterwordspacing}{\spaceskip=0pt\relax}
\providecommand{\BIBentryALTinterwordstretchfactor}{4}
\providecommand{\BIBentryALTinterwordspacing}{\spaceskip=\fontdimen2\font plus
\BIBentryALTinterwordstretchfactor\fontdimen3\font minus \fontdimen4\font\relax}
\providecommand{\BIBforeignlanguage}[2]{{%
\expandafter\ifx\csname l@#1\endcsname\relax
\typeout{** WARNING: IEEEtran.bst: No hyphenation pattern has been}%
\typeout{** loaded for the language `#1'. Using the pattern for}%
\typeout{** the default language instead.}%
\else
\language=\csname l@#1\endcsname
\fi
#2}}
\providecommand{\BIBdecl}{\relax}
\BIBdecl

\bibitem{radford_robust_2023}
A.~Radford, J.~W. Kim, T.~Xu, G.~Brockman, C.~Mcleavey, and I.~Sutskever, ``\BIBforeignlanguage{en}{Robust {Speech} {Recognition} via {Large}-{Scale} {Weak} {Supervision}},'' in \emph{\BIBforeignlanguage{en}{Proceedings of the 40th {International} {Conference} on {Machine} {Learning}}}.\hskip 1em plus 0.5em minus 0.4em\relax PMLR, Jul. 2023, pp. 28\,492--28\,518, iSSN: 2640-3498.

\bibitem{cifka_lyrics_2024}
O.~Cífka, H.~Schreiber, L.~Miner, and F.-R. Stöter, ``Lyrics {Transcription} for {Humans}: {A} {Readability}-{Aware} {Benchmark},'' in \emph{Proceedings of the 25th {International} {Society} for {Music} {Information} {Retrieval} {Conference}, {ISMIR} 2024, {San} {Francisco}, {California}, {USA} and {Online}, {November} 10-14, 2024}, 2024, pp. 737--744.

\bibitem{demirel_mstre-net_2021}
E.~Demirel, S.~Ahlbäck, and S.~Dixon, ``{MSTRE}-{Net}: {Multistreaming} {Acoustic} {Modeling} for {Automatic} {Lyrics} {Transcription},'' in \emph{Proceedings of the 22nd {International} {Society} for {Music} {Information} {Retrieval} {Conference}, {ISMIR} 2021, {Online}, {November}}, 2021, pp. 151--158.

\bibitem{ou_transfer_2022}
L.~Ou, X.~Gu, and Y.~Wang, ``Transfer {Learning} of wav2vec 2.0 for {Automatic} {Lyric} {Transcription},'' in \emph{Proceedings of the 23rd {International} {Society} for {Music} {Information} {Retrieval} {Conference}, {ISMIR} 2022, {Bengaluru}, {India}, {December} 4-8, 2022}, 2022, pp. 891--899.

\bibitem{zhuo_lyricwhiz_2023}
L.~Zhuo, R.~Yuan, J.~Pan, Y.~Ma, Y.~Li, and G.~Zhang, ``{LyricWhiz}: {Robust} {Multilingual} {Zero}-{Shot} {Lyrics} {Transcription} by {Whispering} to {ChatGPT},'' in \emph{Proceedings of the 24th {International} {Society} for {Music} {Information} {Retrieval} {Conference}, {ISMIR} 2023, {Milan}, {Italy}, {November} 5-9, 2023}, 2023, pp. 343--351.

\bibitem{antonisen_polysinger_2024}
S.~Antonisen and I.~López-Espejo, ``{PolySinger}: {Singing}-{Voice} to {Singing}-{Voice} {Translation} {From} {English} to {Japanese},'' in \emph{Proceedings of the 25th {International} {Society} for {Music} {Information} {Retrieval} {Conference}, {ISMIR} 2024, {San} {Francisco}, {California}, {USA} and {Online}, {November} 10-14, 2024}, 2024, pp. 688--696.

\bibitem{fine_making_2014}
P.~A. Fine and J.~Ginsborg, ``\BIBforeignlanguage{English}{Making myself understood: perceived factors affecting the intelligibility of sung text},'' \emph{\BIBforeignlanguage{English}{Frontiers in Psychology}}, vol.~5, Sep. 2014, publisher: Frontiers.

\bibitem{defossez_hybrid_2021}
\BIBentryALTinterwordspacing
A.~Défossez, ``Hybrid {Spectrogram} and {Waveform} {Source} {Separation},'' \emph{CoRR}, vol. abs/2111.03600, 2021, arXiv: 2111.03600. [Online]. Available: \url{https://arxiv.org/abs/2111.03600}
\BIBentrySTDinterwordspacing

\bibitem{chambers1980nonlexical}
\BIBentryALTinterwordspacing
C.~K. Chambers, ``Non-lexical vocables in scottish traditional music,'' PhD thesis, University of Edinburgh, 1980, accessed: May 14, 2025. [Online]. Available: \url{https://era.ed.ac.uk/handle/1842/7220}
\BIBentrySTDinterwordspacing

\bibitem{romana_automatic_2024}
A.~Romana, K.~Koishida, and E.~M. Provost, ``Automatic {Disfluency} {Detection} {From} {Untranscribed} {Speech},'' \emph{IEEE/ACM Transactions on Audio, Speech, and Language Processing}, vol.~32, pp. 4727--4740, 2024.

\bibitem{stoller_end--end_2019}
D.~Stoller, S.~Durand, and S.~Ewert, ``End-to-end {Lyrics} {Alignment} for {Polyphonic} {Music} {Using} an {Audio}-to-character {Recognition} {Model},'' in \emph{{ICASSP} 2019 - {IEEE} {International} {Conference} on {Acoustics}, {Speech} and {Signal} {Processing}}, May 2019, pp. 181--185, iSSN: 2379-190X.

\bibitem{meseguer-brocal_dali_2018}
G.~Meseguer-Brocal, A.~Cohen-Hadria, and G.~Peeters, ``{DALI}: {A} {Large} {Dataset} of {Synchronized} {Audio}, {Lyrics} and notes, {Automatically} {Created} using {Teacher}-student {Machine} {Learning} {Paradigm}.'' in \emph{Proceedings of the 19th {International} {Society} for {Music} {Information} {Retrieval} {Conference}, {ISMIR} 2018, {Paris}, {France}, {September}}, 2018, pp. 431--437.

\bibitem{schulze-forster_phoneme_2021}
K.~Schulze-Forster, C.~S.~J. Doire, G.~Richard, and R.~Badeau, ``\BIBforeignlanguage{en}{Phoneme {Level} {Lyrics} {Alignment} and {Text}-{Informed} {Singing} {Voice} {Separation}},'' \emph{\BIBforeignlanguage{en}{IEEE/ACM Transactions on Audio, Speech and Language Proc.}}, 2021.

\bibitem{rafii_musdb18-hq_2019}
\BIBentryALTinterwordspacing
Z.~Rafii, A.~Liutkus, F.-R. Stöter, S.~I. Mimilakis, and R.~Bittner, ``{MUSDB18}-{HQ} - an uncompressed version of {MUSDB18},'' Aug. 2019. [Online]. Available: \url{https://zenodo.org/records/3338373}
\BIBentrySTDinterwordspacing

\bibitem{bain_whisperx_2023}
M.~Bain, J.~Huh, T.~Han, and A.~Zisserman, ``{WhisperX}: {Time}-{Accurate} {Speech} {Transcription} of {Long}-{Form} {Audio},'' in \emph{24th {Annual} {Conference} of the {International} {Speech} {Communication} {Association}, {Interspeech} 2023, {Dublin}, {Ireland}, {August}}.\hskip 1em plus 0.5em minus 0.4em\relax ISCA, 2023, pp. 4489--4493.

\bibitem{ye_spurious_2024}
\BIBentryALTinterwordspacing
W.~Ye, G.~Zheng, X.~Cao, Y.~Ma, and A.~Zhang, ``Spurious {Correlations} in {Machine} {Learning}: {A} {Survey},'' May 2024, arXiv:2402.12715 [cs]. [Online]. Available: \url{http://arxiv.org/abs/2402.12715}
\BIBentrySTDinterwordspacing

\bibitem{bonzi_exploiting_2023}
F.~Bonzi, M.~Mancusi, S.~D. Deo, P.~Melucci, M.~S. Tavella, and L.~Parisi, ``Exploiting {Music} {Source} {Separation} {For} {Singing} {Voice} {Detection},'' in \emph{{IEEE} 33rd {International} {Workshop} on {Machine} {Learning} for {Signal} {Processing} ({MLSP})}, Sep. 2023, pp. 1--6, iSSN: 2161-0371.

\bibitem{zusag24_interspeech}
M.~Zusag, L.~Wagner, and B.~Thallinger, ``Crisperwhisper: Accurate timestamps on verbatim speech transcriptions,'' in \emph{Interspeech 2024}, 2024, pp. 1265--1269.

\end{thebibliography}

\end{document}